\let\LATEXcline\cline
\documentclass[sn-mathphys-num]{sn-jnl} 
\let\cline\LATEXcline

\usepackage[english]{babel}
\usepackage[title]{appendix}%
\usepackage{booktabs}%
\usepackage{amsmath, amssymb, mathtools, physics}
\usepackage{graphicx,float}
\usepackage{multirow}
\usepackage[RGB,dvipsnames]{xcolor}
\usepackage{academicons}
\definecolor{dodgerblue}{rgb}{0.12, 0.56, 1.0}
\definecolor{darkcyan32144140}{RGB}{32,144,140}         %
\definecolor{darkslateblue5294141}{RGB}{52,94,141}      
\definecolor{darkslateblue6467135}{RGB}{64,67,135}      
\definecolor{darkslateblue7235116}{RGB}{72,35,116}      
\definecolor{darkslategray38}{RGB}{38,38,38}
\definecolor{greenyellow18922238}{RGB}{189,222,38}      
\definecolor{indigo68184}{RGB}{68,1,84}                 
\definecolor{lavender234234242}{RGB}{234,234,242}
\definecolor{mediumseagreen34167132}{RGB}{34,167,132}   
\definecolor{mediumseagreen68190112}{RGB}{68,190,112}   %
\definecolor{teal41120142}{RGB}{41,120,142}
\definecolor{yellowgreen12120981}{RGB}{121,209,81}
\newcommand{\snippet}[4]{\gamma_{#1\to #2}(#3, #4)}
\newcommand{\Snippet}[4]{\Gamma_{#1\to #2}(#3, #4)}
\newcommand{\wtdpsiNESS}[3]{\psi_{#1\to #2}(#3)}
\newcommand{\wtdpsi}[4]{\psi_{#1\to #2}(#3\vert #4)}
\newcommand{\wtdpsitrev}[4]{\tilde{\psi}_{#1\to #2}(#3\vert #4)}
\newcommand{\wtdPsi}[3]{\Psi_{#1\to #2}(#3)}

\newcommand{\meanness}[1]{\left\langle #1\right\rangle_\text{ss}}
\newcommand{\meanpss}[1]{\left\langle #1\right\rangle_\text{pss}}
\newcommand{\psischaetz}{\left\langle \hat{\sigma}_\psi\right\rangle_\text{pss}}
\newcommand{\Psischaetz}{\left\langle \hat{\sigma}_\Psi\right\rangle_\text{pss}}
\newcommand{\pkschaetz}{\left\langle \hat{\sigma}_{pk}\right\rangle_\text{pss}}
\newcommand{\pathw}[1]{\mathcal{P}\left[#1\right]}
\newcommand{\pathwrev}[1]{\widetilde{\mathcal{P}}\left[#1\right]}
\usepackage{hyperref}
\usepackage{tikz}
\usetikzlibrary{external}
\tikzsetexternalprefix{tikz/}
\usepackage{pgfplots}
\usetikzlibrary{automata, arrows.meta, positioning, calc, decorations.pathreplacing, calligraphy, backgrounds}
\begin{document}
%
\title[Inferring kinetics and entropy production from observable transitions in partially accessible, periodically driven Markov networks]{Inferring kinetics and entropy production from observable transitions in partially accessible, periodically driven Markov networks}
\author[1]{\fnm{Alexander M.} \sur{Maier}}\email{amaier@theo2.physik.uni-stuttgart.de}
\author[1]{\fnm{Julius} \sur{Degünther}}\email{deguenther@theo2.physik.uni-stuttgart.de}
\author[1]{\fnm{Jann} \spfx{van der} \sur{Meer}}\email{vdmeer@theo2.physik.uni-stuttgart.de}
\author*[1]{\fnm{Udo} \sur{Seifert}}\email{useifert@theo2.physik.uni-stuttgart.de}
\affil[1]{\orgdiv{II. Institut für Theoretische Physik}, \orgname{Universität Stuttgart}, \orgaddress{\street{Pfaffenwaldring 57}, \city{Stuttgart}, \postcode{70550}, \country{Germany}}} 
\date{\today} 
\abstract{For a network of discrete states with a periodically driven Markovian dynamics, we develop an inference scheme for
an external observer who has access to some transitions. Based on waiting-time distributions between these transitions, the
periodic probabilities of states connected by these observed transitions and their time-dependent transition rates
can be inferred. Moreover, the smallest number of hidden transitions between accessible ones and some of
their transition rates can be extracted. We prove and conjecture lower bounds on the total entropy production for such
periodic stationary states. Even though our techniques are based on generalizations of known methods for steady states, we
obtain original results for those as well.}

%
\maketitle
\section{Introduction}
The framework of stochastic thermodynamics provides rules to describe small physical systems 
that are embedded into a thermal reservoir but remain out of equilibrium due to external 
driving \cite{seif12, peli21, shir23}. If the relevant degreees of freedom can be described by a memoryless, i.e., Markovian
dynamics on a discrete set of states, the time-evolution of the system is governed by the
network structure and the transition rates between the states. In the case of periodically driven 
transition rates, such a dynamics relaxes into a periodic stationary state (PSS) \cite{raha08,cher08,raz16a,rots16}, which, as a 
special case, becomes a non-equilibrium steady state (NESS) \cite{stig11,rold10,muy13,ge12} for constant transition rates. 

Since a model is fully specified only if all transition rates are known, practically relevant 
scenarios in which parts of the model remain hidden \cite{espo12,arig18} require methods to recover, e.g., hidden
transition rates on the basis of observable data of a particular form. The combination of 
such methods with the physical constraints provided by the rules of stochastic
thermodynamics comprises the field of thermodynamic inference \cite{seif19}. With a focus on
quantities that have a thermodynamic interpretation, recent works in the field obtain bounds 
on entropy production \cite{mart19, dech21, skin21, haru22, vdm22, vdm22b} or affinities 
\cite{vdm22, ohga23, lian23, degu23}, which
are complemented with techniques to recover topological information \cite{li13, vu23} and speed limits
\cite{ito20, shir18, dech23}.

Many of the methods discussed above apply to the case of time-independent driving and cannot
straightforwardly be generalized to a PSS. For one of the standard methods of estimating entropy production,
the thermodynamic uncertainty relation \cite{bara15, ging16}, generalizations to PSSs exist
\cite{proe17, bara18b, koyu19a, bara19}, which require more input than their time-independent counterparts in general.

For the purpose of estimating entropy production, the usual rationale if given information
about residence in states is to identify appropriate transitions or 
currents, since such time-antisymmetric data allow one to infer the entropy production. When observing 
transitions, one can ask the converse question: Can we infer information
about states, which are time-symmetric, from antisymmetric data like transitions? We will address in this work how
observing transitions allows us to recover occupation probabilites in states if the system
is in a PSS. In addition, we will generalize and extend methods from \cite{vdm22} to the 
periodically driven case to infer transition rates and the number of hidden transitions 
between two observable ones.
We also formulate and compare different lower bounds on the 
mean total entropy production. These entropy estimators are either proved or supported 
with strong numerical evidence. 

The paper is structured as follows. In Section \ref{sec:generalSetup}, we describe the setting and identify
waiting-time distributions between observed transitions as the basic quantities we use 
to formulate our results. In Section \ref{sec:infkinetics}, we investigate how these quantities can be used 
to infer kinetic information about the hidden part of a system in a PSS or NESS.
Estimators for the mean entropy production are discussed in Section \ref{sec:ALLestimators}. We conclude
and give an outlook on further work in Section \ref{sec:conclusion}.
\section{General setup}\label{sec:generalSetup}
We consider a network of $N$ states $i\in\left\lbrace 1,\dots,N\right\rbrace$ that is periodically driven. The system is
in state $i(t)$ at time $t$ and follows a stochastic description by allowing transitions between states sharing an edge in
the graph. A transition from $k$ to $l$ happens instantaneously with rate $k_{kl}(t)$, which has the periodicity of the
driving. To ensure thermodynamic consistency, we assume the local detailed balance condition \cite{seif12, peli21, shir23}
\begin{align}
\frac{k_{kl}(t)}{k_{lk}(t)} = e^{F_{k}(t) - F_{l}(t) + f_{kl}(t)}
\label{eq:localDetailedBalance}
\end{align}
at each link, i.e., for each transition and its reverse.
The driving with period $\mathcal{T}$ may change the free energy $F_k(t)$ of states $k$ or act as a non-conservative
force along transitions from $k$ to $l$ with $f_{kl}(t) = -f_{lk}(t)$. Energies in this work are given in units of
thermal energy so that entropy production is dimensionless. \par
The dynamics of the probability $p_k(t)$ to occupy state $k$ at time $t$ obeys the master equation
\begin{align}
\partial_t p_k(t) = \sum_{l} \left[-p_k(t)k_{kl}(t) + p_l(t)k_{lk}(t)\right]. \label{eq:mastereq}
\end{align}
In the long-time limit $t\to\infty$, these networks approach a periodic stationary state (PSS) $p^\text{pss}_k(t)$.
The transition rates and these probabilities $p^\text{pss}_k(t)$ determine the mean entropy production rate in the PSS
\cite{seif12, peli21, shir23}
\begin{align}
\meanpss{\sigma} \equiv \frac{1}{\mathcal{T}}\int_{0}^{\mathcal{T}} \sum_{kl}p^\text{pss}_k(t)k_{kl}(t)
\ln\frac{p^\text{pss}_k(t)k_{kl}(t)}{p^\text{pss}_l(t)k_{lk}(t)} \dd{t}. \label{eq:setupentropysigma}
\end{align} \par
\begin{figure}
\centering
\includegraphics[scale=1]{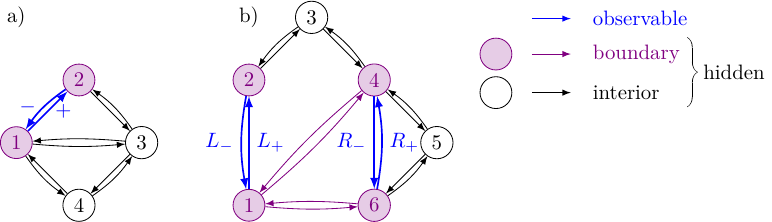}
\caption{Graphs of partially observable, periodically driven Markov networks. Observable transitions are labeled and
displayed in blue. All states and the remaining transitions are assumed to be hidden. Purple states and purple transitions
form the boundary of the hidden network. In a), only one pair of transitions can be observed. In b), the two pairs
$1\leftrightarrow 2$ and $4\leftrightarrow 6$ are observable. The whole network consists of the observable transitions, the
boundary of the hidden network and its interior.}
\label{fig:graphExamples}
\end{figure}

In this work, we assume that at least one pair of transitions of a Markov network in its PSS or NESS is observable for an
external observer while other transitions and all states are hidden, i.e., not directly accessible for the observer. We
illustrate this with graphs of two exemplary Markov networks in Figure \ref{fig:graphExamples}. States with an observable
transition between them will be called boundary states. If two boundary states are connected with one hidden transition,
these transitions and the boundary states form the boundary of the hidden network. Additionally, we assume the period
$\mathcal{T}$ of the driving to be known. \par
The task is to determine hidden quantities like the probabilities $p^\text{pss}_k(t)$ of such partially accessible networks
as well as to estimate the overall entropy production.
In such a network, we can determine distributions of waiting times $t$ between two successive observable transitions
$I=(ij)$ and $J=(kl)$, whereas observing the full microscopic dynamics is impossible. These waiting-time distributions
are of the form
\begin{align}
\wtdpsi{I}{J}{t}{t_0} &\equiv \sum_{\snippet{I}{J}{t}{t_0}} \mathcal{P}\left[\snippet{I}{J}{t}{t_0}
\middle\vert I,t_0\right].
\label{eq:wtdDefpsismall}
\end{align}
They depend on the time $t_0\in[0,\mathcal{T}]$ at which transition $I$ occurs within one period of the PSS. Since an
arbitrary number of hidden transitions occurs between $I$ and $J$, the distributions are given by the sum of conditional
path weights $\mathcal{P}\left[\snippet{I}{J}{t}{t_0} \middle\vert I,t_0\right]$ corresponding to all microscopic
trajectories $\snippet{I}{J}{t}{t_0}$ that start directly after a transition $I$ at $t_0$ and end with the next observable
transition $J$ after waiting time $t$. \par
Furthermore, we define
\begin{align}
\wtdPsi{I}{J}{t} = \int_{0}^{\mathcal{T}} p^\text{pss}(t_0\vert I)\wtdpsi{I}{J}{t}{t_0}\dd{t_0},
\label{eq:wtdDefPsicapital}
\end{align}
where we use the conditional probability $p^\text{pss}(t_0\vert I)$ to detect a particular transition $I$ at a specific time
$t_0\in[0,\mathcal{T})$ within the period. Due to effectively marginalizing $t_0$ like in Equation
\eqref{eq:wtdDefPsicapital} when using trajectories with uncorrelated $t_0$, e.g., observed trajectories for unknown
$\mathcal{T}$ in which we discard a sufficient number of successive waiting times between two saved ones, we can always get
these waiting-time distributions from measured waiting times. In the special case of a NESS,
\begin{align}
\wtdpsi{I}{J}{t}{t_0} = \wtdPsi{I}{J}{t} \equiv \wtdpsiNESS{I}{J}{t} \label{eq:setupwtdinNESS}
\end{align}
holds for an arbitrarily assigned period $\mathcal{T}$, which we emphasize by using $\wtdpsiNESS{I}{J}{t}$.
\section{Shortest hidden paths, transition rates and occupation probabilities}\label{sec:infkinetics}
We first generalize methods to infer the number of hidden transitions in the shortest path between any two observable
transitions from a NESS \cite{vdm22, li13} to a PSS. For any two transitions $I,J$ for which the waiting-time distribution
does not vanish, the number of hidden transitions $M_{IJ}$ along the shortest path between $I$ and $J$ is given by
\begin{align}
M_{IJ} &= \lim_{t\to 0} \left(t \dv{t}\ln\left[\wtdpsi{I}{J}{t}{t_0}\right] \right)
= \lim_{t\to 0} \left(t \dv{t}\ln\left[\wtdPsi{I}{J}{t}\right] \right), \label{eq:topoN1}
\end{align}
which can be derived following an idea adopted in reference \cite{vdm22} for systems in a NESS. For systems in a PSS, the
waiting-time distributions $\wtdpsi{I}{J}{t}{t_0}$ and $\wtdPsi{I}{J}{t}$ are interchangeable since both of their short-time
limits are proportional to $t^{M_{IJ}}$, i.e., are dominated by the shortest path between $I$ and $J$. Since this shortest
path between $I$ and $J$ consists of the same number of transitions, no matter at what time $t_0$ a trajectory starts, the
expression in the middle of Equation \eqref{eq:topoN1} is independent of $t_0$. In the example of Figure
\ref{fig:graphExamples}\,a), with $I=+$ and $J=+$, we get $M_{++}=2$ since the two transitions $(23)$ and $(31)$ form the
corresponding shortest hidden path. For the graph shown in Figure \ref{fig:graphExamples}\,b) with $I=L_-$ and $J=R_+$, we
find $M_{L_-R_+}=1$ since $(16)$ is the transition inbetween $L_-$ and $R_+$. \par
The rates of observable transitions can be recovered from waiting-time distributions. Given a transition $J=(kl)$ and its
reverse $\tilde{J}=(lk)$, we obtain the corresponding rate $k_{kl}$ through
\begin{align}
\lim_{t\to 0} \wtdpsi{\tilde{J}}{J}{t}{t_0} &= \lim_{t\to 0}k_{kl}(t+t_0)p_k(t\vert\tilde{J},t_0)
= k_{kl}(t_0), \label{eq:getkijVIS}
\end{align}
where we use that the short-time limit is dominated by the path with the fewest transitions, which, in this case, is
$\tilde{J}$ followed by $J$ without any hidden intermediate transitions. Moreover, the state of the system is known
immediately after the initial transition $\tilde{J}$ at time $t_0$ within the period, which leads to
$p_k(t=0\vert\tilde{J},t_0)=1$. \par
Further transition rates are inferable for any combination of transitions $I$ and $J$ with $M_{IJ} = 1$, i.e., whenever the
shortest path between $I=(ij)$ and $J=(kl)$ consists of only one hidden transition. For $t\to 0$, its transition rate
$k_{jk}(t_0)$ then follows from a Taylor expansion as
\begin{align}
k_{jk}(t_0) = \lim_{t\to 0} \frac{\wtdpsi{I}{J}{t}{t_0}}{k_{kl}(t_0+t)t}
\label{eq:getkijTaylor}
\end{align}
with $k_{kj}(t_0)$ following analogously. As an example, we get the transition rates $k_{14},\,k_{16},\,k_{41}$ and $k_{61}$
for the network shown in Figure \ref{fig:graphExamples}\,b) even though the related links are hidden. \par
Occupation probabilities of boundary states of the hidden network can be inferred as follows. During a measurement of length
$M\mathcal{T}$ with large $M\in\mathbb{N}$, we count the number $N_I(t_0\leq\tau\leq t_0+\Delta t)$ of transitions $I=(ij)$
that occur during the infinitesimal interval $[t_0,t_0+\Delta t]$, where we map all times at which transitions happen into
one period of the PSS using a modulo operation. We, therefore, obtain the rate of transitions $I$ at time
$t_0\in[0,\mathcal{T})$ within one period of the PSS as
\begin{align}
n_I(t_0) = \lim_{M\to\infty}\lim_{\Delta t\to 0} \frac{N_I(t_0\leq\tau\leq t_0+\Delta t)}{M\Delta t}
= p_i^\text{pss}(t_0)k_{ij}(t_0). \label{eq:pIpssnItnullone}
\end{align}
As the transition rate $k_{ij}(t_0)$ can be determined as described above, we can thus infer $p_i^\text{pss}(t_0)$ from
experimentally accessible data. Knowing the occupation probabilities of all boundary states of the hidden network allows us
to calculate instantaneous currents along single transitions between them using the corresponding inferred transition
rates. \par
These results can be specialized to NESSs, where, to the best of our knowledge, they have not been reported yet either. In
this special case, dropping the irrelevant $t_0$ in Equations \eqref{eq:getkijVIS} and \eqref{eq:getkijTaylor} leads to
constant transition rates. Moreover, in a NESS, the mean rate of transitions $I$, $\meanness{n_I}$, can directly be obtained
from the total number $N_{I,T}$ of observed transitions $I$ along a measured trajectory of length $T$. Inferring occupation
probabilities $p_i^\text{ss}$ then only requires dividing through the already calculated transition rate $k_{ij}$, i.e.,
\begin{align}
p_i^\text{ss}k_{ij} = \meanness{n_I} = \lim_{T\to\infty}\frac{N_{I,T}}{T}. \label{eq:pIssnINESS}
\end{align} \par
\begin{figure}
\centering
\hspace{-.5em}\includegraphics[scale=1]{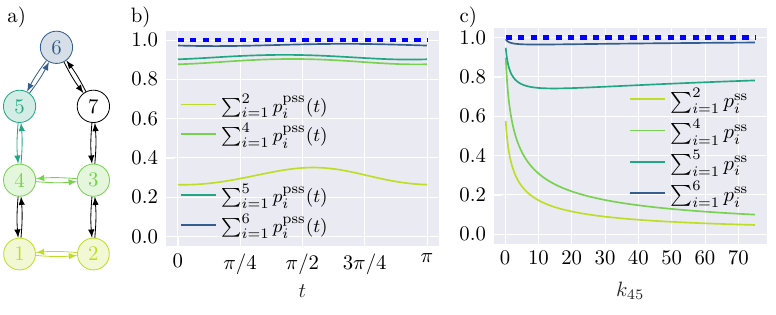}\hspace{-1em}
\caption{Inference of occupation probabilities of boundary states. a) Network with hidden states and hidden (black)
transitions. Beginning with the light green transitions $(12)$ and $(21)$ in a), darker colored transitions are successively
considered as observable. b) The sums of occupation probabilities of states within the boundary of each hidden network in
the PSS are shown in the respective color. The time-dependent transition rates are given in Appendix
\ref{appsec:mingraphsimparams}. c) As in b) but for constant driving, i.e., for a NESS as a function of $k_{45}$ with the
other rates given in Appendix \ref{appsec:mingraphsimparams}.}
\label{fig:migsumofppss}
\end{figure}
Through equations \eqref{eq:pIpssnItnullone} and \eqref{eq:pIssnINESS} we show how to infer occupation probabilities of
boundary states of the hidden network. Given the inferable quantities $p_i^\text{pss}(t_0)$ or $p_i^\text{ss}$, we can
calculate how much probability rests on states in the network beyond the so identified boundary states. As an example,
Figures \ref{fig:migsumofppss}\,b) and c) illustrate the probability to find the network shown in Figure
\ref{fig:migsumofppss}\,a) in its boundary states rather than in states within the interior of the hidden network. In both
figures, different sets of observable transitions lead to different boundaries of the hidden network. Figure
\ref{fig:migsumofppss}\,b) displays sums of probabilities for these systems in a PSS, while Figure
\ref{fig:migsumofppss}\,c) gives an example for NESSs. Each sum of inferable occupation probabilities quantifies the
probability of finding the system in the boundary of the hidden network. The closer this sum is to one, the less relevant
the inaccessible states in the interior of the hidden network are for the dynamics. \par
\section{Three estimators for entropy production in PSS\lowercase{s}}\label{sec:ALLestimators}
In this section, we estimate irreversibility via the entropy production rate in a PSS. We have seen above how waiting-time
distributions contain information on the hidden dynamics of a network. Thus, it seems sensible to expect that these
quantities can be used as entropy estimator to infer irreversibility in both the observable and the hidden parts of the
network. \par
For a trajectory $\Gamma$ of length $T$, reversing the driving protocol leads to transition rates $\tilde{k}_{ij}(t)=
k_{ij}(T-t)$. The corresponding waiting-time distributions $\wtdpsitrev{\tilde{J}}{\tilde{I}}{t}{t_0+t}$ for reversed paths
$\tilde{J}\to\tilde{I}$ are the time-reversed versions of $\wtdpsi{I}{J}{t}{t_0}$. Once waiting-time distributions of the
form $\wtdpsitrev{\tilde{J}}{\tilde{I}}{t}{t_0+t}$ have been determined, the fluctuation relation
\begin{align}
\hat{\sigma}_\psi \equiv \lim_{T\to\infty}\frac{1}{T} \ln\frac{\mathcal{P}\left[\Gamma\right]}{
\widetilde{\mathcal{P}}[\widetilde{\Gamma}]}
\label{eq:estimatorFT}
\end{align}
for a trajectory $\Gamma$ of length $T$ and its time-reverse $\widetilde{\Gamma}$ allows us to derive an estimator
$\psischaetz$ that fulfills
\begin{align}
\meanpss{\sigma} \geq \psischaetz =\sum_{I,J} \int_{0}^{\infty}\int_{0}^{\mathcal{T}} \frac{n_I(t_0)}{\mathcal{T}}
\wtdpsi{I}{J}{t}{t_0} \ln\frac{\wtdpsi{I}{J}{t}{t_0}}{\wtdpsitrev{\tilde{J}}{\tilde{I}}{t}{t_0+t}}
\dd{t_0}\dd{t} \geq 0.
\label{psiseq:psisEPRPsiEstim}
\end{align}
Here, the index $\psi$ of the estimator highlights the type of waiting-time distribution that enters its expression in the
above inequality. We prove this inequality in Appendix \ref{appen:estimatorpsi} as a generalization of the trajectory-based
entropy estimator $\left\langle \hat{\sigma}\right\rangle$ introduced in \cite{vdm22} for NESSs. \par
For time-symmetric driving, estimating $\meanpss{\sigma}$ with $\psischaetz$ does not require to reverse the driving
protocol in an experiment. In this case, $\wtdpsitrev{\tilde{J}}{\tilde{I}}{t}{t_0+t}$ results from $\wtdpsi{I}{J}{t}{t_0}$
by exploiting the symmetry $k_{ij}(t_*+t_0)=k_{ij}(t_*-t_0)$ of the protocol for all transitions $(ij)$ after finding
$t_*\in[0,\mathcal{T})$. In the next paragraphs, we discuss experimentally accessible entropy estimators that do not require
waiting-time distributions of the time-reversed process regardless of symmetry properties of the driving. \par
\begin{figure}
\centering
\hspace{-.95em}\includegraphics[scale=1]{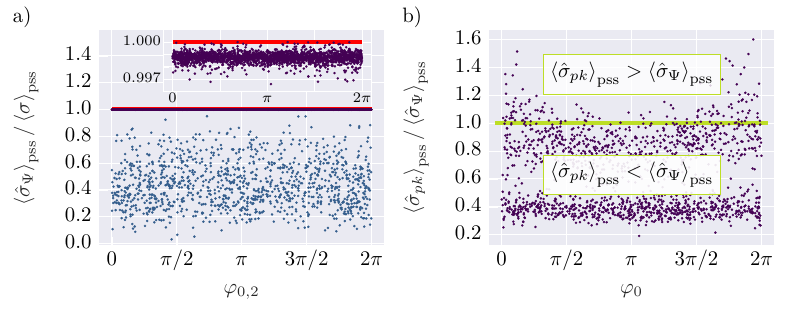}\hspace{-1.05em}
\caption{Ratios $\meanpss{\sigma}/\Psischaetz$ and $\pkschaetz/\Psischaetz$ involving entropy estimators in scatter plots.
a) Quality factor $\Psischaetz/\meanpss{\sigma}$ for two data sets of networks with diamond-shaped graph as shown in Figure
\ref{fig:graphExamples}\,a). b) Comparison between $\pkschaetz$ and $\Psischaetz$ for unicyclic three-state systems. The
ratios in both scatter plots are plotted against the random angle $\varphi_0$ that is part of the free energy
parametrization as detailed in Appendix \ref{appen:paramsconjvgl}.}
\label{fig:conjecture_vgl}
\end{figure}

We have performed extensive numerical computations of random, periodically driven Markov networks corresponding to different
underlying graphs to compute
\begin{align}
\Psischaetz \equiv \sum_{I,J} \meanpss{n_I}
\int_{0}^{\infty} \wtdPsi{I}{J}{t} \ln\frac{\wtdPsi{I}{J}{t}}{\wtdPsi{\tilde{J}}{\tilde{I}}{t}} \dd{t},
\label{psiseq:PsisEPRPsiEstimconj}
\end{align}
where the index $\Psi$ indicates the type of waiting-time distribution used. Here, $\meanpss{n_I}$ is the mean of $n_I(t_0)$
in one period $t_0\in[0,\mathcal{T}]$ that results from measured data. For over $10^5$ randomly chosen systems from
unicyclic graphs of three states, diamond-shaped graphs as displayed in Figure \ref{fig:graphExamples}\,a) and more complex
underlying graphs, the inequalities
\begin{align}
\meanpss{\sigma}\geq \Psischaetz \geq 0
\label{psiseq:Psiconjineq}
\end{align}
hold true as shown in the scatter plot in Figure \ref{fig:conjecture_vgl}\,a). Therefore, we conjecture inequality
\eqref{psiseq:Psiconjineq} to hold true for periodically driven Markov networks, so that $\Psischaetz$ is a
thermodynamically consistent estimator of $\meanpss{\sigma}$. \par
Furthermore, transition rates and occupation probabilities that are inferred as described in Section \ref{sec:infkinetics}
allow us to prove another lower bound on the entropy production rate of a Markov network in a PSS which complements the
previous two. The estimator
\begin{align}
\pkschaetz \equiv \int_{0}^{\mathcal{T}} \sum_{ij\in\mathcal{V}} \frac{p^\text{pss}_i(t) k_{ij}(t)}
{\mathcal{T}} \ln\frac{p^\text{pss}_i(t) k_{ij}(t)}{p^\text{pss}_j(t) k_{ji}(t)} \dd{t}
\label{eq:estimatorpk}
\end{align}
adds up the contributions to entropy production along transitions of the set $\mathcal{V}$ containing all transitions that
are either observable or within the boundary of the hidden network. As $\pkschaetz$ solely depends on inferable
probabilities and rates, its index is $pk$. Since each of the terms in Equation \eqref{eq:estimatorpk} is non-negative for
all $t$ and part of $\meanpss{\sigma}$ as given in Equation \eqref{eq:setupentropysigma}, $\pkschaetz$ constitutes a lower
bound on the total entropy production rate of the system. This bound may often be less tight than the conjectured bound
\eqref{psiseq:PsisEPRPsiEstimconj} for periodically driven Markov networks though this ordering does not hold in general as
shown in Figure \ref{fig:conjecture_vgl}\,b). \par
In the special case of a NESS, the last bound \eqref{eq:estimatorpk} acquires the familiar form
\begin{align}
\pkschaetz \overset{\text{NESS}}{=} \sum_{ij\in\mathcal{V}}
p^\text{ss}_i k_{ij} \ln\frac{p^\text{ss}_i k_{ij}}{p^\text{ss}_j k_{ji}}.
\label{eq:estimatorpkNESS}
\end{align}
The crucial part is that here the entropy estimator $\pkschaetz$ is based on occupation probabilities and transition rates
inferred from distributions of waiting times between observable transitions as described in Section \ref{sec:infkinetics}.
Although the term \eqref{eq:estimatorpkNESS} shows superficial similarities to the main result of reference \cite{shir14}
interpreted as an entropy estimator, our estimator $\pkschaetz$ differs in two ways. First, $\pkschaetz$ can be used for
partially accessible Markov networks in a PSS, which include systems in a NESS as special cases. Second, the sum in
Equations \eqref{eq:estimatorpk} and \eqref{eq:estimatorpkNESS} includes contributions of both observable transitions and
transitions within the boundary of the hidden network. These additional contributions allow for a more accurate estimate of
the entropy production rate.
\section{Concluding perspective}\label{sec:conclusion}
\begin{figure}
\centering
\includegraphics[scale=1]{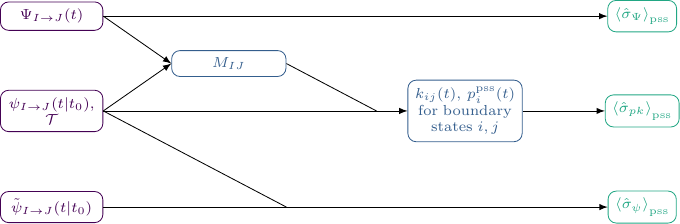}
\caption{Summary of the inference scheme. Starting from waiting-time distributions and the known period $\mathcal{T}$ of the
driving, the number of hidden transitions $M_{IJ}$ along the shortest path between two observable transitions $I$ and $J$,
the occupation probabilities of boundary states $p_i^\text{pss}(t)$ as well as the rates $k_{ij}(t)$ of transitions between
boundary states are inferable. These quantities enter the three lower bounds on entropy production.}
\label{fig:redDepGraph}
\end{figure}

In this paper, we have introduced inference methods based on distributions of waiting times between consecutive observed
transitions in partially accessible, periodically driven Markov networks. Successive use of these methods yields information
about the kinetics of such a Markov network as well as its underlying topology, including hidden parts, as summarized in
Figure \ref{fig:redDepGraph}. \par
We have first shown how to infer the number of hidden transitions along the shortest path between two observable
transitions. We have then derived methods to infer transition rates between boundary states of the hidden network.
Occupation probabilities of these boundary states then follow by discerning when the observable transitions happen within
one period. Consequently, we find the total probability resting on the hidden states in the interior of the hidden network.
\par
In addition, we have presented three entropy estimators enabling us to estimate irreversibility of a driven Markov network
based on observed transitions during a partially accessible dynamics. The first and third one are proven to be lower bounds
of the mean entropy production rate, whereas we conjecture the second estimator to have this property too. Its proof remains
as open theoretical challenge. The second and third estimator have the advantage of not requiring control of the driving
since its time reversal is not needed. Furthermore, we emphasize that even for the simpler NESS most of these results are
original as well. \par
Finally, it will be interesting to explore whether and how such an approach can be adapted to continuous systems described
by a Langevin dynamics. We also hope that our non-invasive method yielding time-dependent transition rates and occupation
probabilities will be applied to experimental data of periodically driven small systems. \par
%
%
%
%
%
%
%
%
%
%
%
%
%
%
%
%
%
%
%
\bibliography{references.bib} 

\begin{appendices}
\section{Parameters used for numerical data}
\subsection{Parameters for data shown in Figure \ref{fig:migsumofppss}}\label{appsec:mingraphsimparams}
For the network shown in Figure \ref{fig:migsumofppss}\,a), the PSS is generated with transition rates
\begin{align}
k_{ij}(t) = \kappa_{ij}e^{\left(F_i(t) - F_j(t) + f_{ij}\right)/2} & & \text{and} & & 
k_{ji}(t) = \kappa_{ij}e^{-\left(F_i(t) - F_j(t) + f_{ij}\right)/2}.
\label{Appeq:mgraphtransrates}
\end{align}
Therein, we set all $\kappa_{ij}=1$ as well as $f_{12}=f_{23}=f_{34}=f_{41}=2$ and $f_{45}=f_{56}=f_{67}=f_{73}=20/3$.
Furthermore, we choose the free energies
\begin{align}
F_1(t) &= 1.0 + 0.3\cos{2t}, & & & F_2(t) &= 1.9 + 0.5\sin{2t},\notag \\
F_3(t) &= 1.4 + 0.4\sin{2t}, & & & F_4(t) &= 1.0 + 0.7\sin{2t},\notag \\
F_5(t) &= 3.1 + 0.1\sin{2t}, & & & F_6(t) &= 1.8 + 0.3\sin{2t},\notag \\
F_7(t) &= 2.5
\end{align}
and solve the master equation \eqref{eq:mastereq} of the network for the occupation probabilities $p_i^\text{pss}(t)$. \par
The NESSs for this network as shown in Figure \ref{fig:migsumofppss}\,c) are generated with the non-zero transition
rates $k_{12} = 1.7,\ k_{14} = 0.4,\ k_{21} = 0.6,\ k_{23} = 3.5,\ k_{32} = 0.3,\ k_{34} = 3.3,\ k_{37} = 0.02,\ k_{41}
= 5.7,\ k_{43} = 0.3,\ k_{54} = 0.1,\ k_{56} = 0.7,\ k_{65} = 0.2,\ k_{67} = 0.8,\ k_{73} = 4.6,\ k_{76} = 0.05$ and
$k_{45} \in [0.2,75]$.
\subsection{Parameters for data shown in Figure \ref{fig:conjecture_vgl}}\label{appen:paramsconjvgl}
For Figure \ref{fig:conjecture_vgl}\,a), we have used diamond-shaped networks as shown in Figure
\ref{fig:graphExamples}\,a) but with observable transitions $1\leftrightarrow 2$ and $1\leftrightarrow 4$. For Figure
\ref{fig:conjecture_vgl}\,b), we have used unicyclic three-state systems with observable transitions $1\leftrightarrow 2$.
All transition rates are parameterized as in Equation \eqref{Appeq:mgraphtransrates} with $\kappa_{ij}=1$ unless otherwise
specified. \par
All diamond-shaped systems are characterized by $f_{12}=f_{14}=f_{23}=f_{31}=f_{43}=2$. The free energies of the states in
each simulated diamond network are given by
\begin{align}
F_1(t) &= F_{c1} + F_{a1}\sin{\omega t} \\
F_2(t) &= F_{c2} + F_{a2}\sin(n_{\omega,2}\omega t+ \varphi_{0,2}) \\
F_3(t) &= F_{c3} + F_{a3}\sin(n_{\omega,3}\omega t+ \varphi_{0,3}) \\
F_4(t) &= F_{c4} + F_{a4}\sin(n_{\omega,4}\omega t+ \varphi_{0,4}),
\end{align}
where constant energies $F_{ci}$, energy amplitudes $F_{ai}$ and the angles $\varphi_{0,i}$ are randomly picked from
normal distributions with mean and variance as given in Table \ref{tab:conjParamRanges}.
\begin{table}
\centering
\caption{Mean and variance of the normal distribution corresponding to the two parameter sets,
\textcolor{indigo68184}{indigo} and \textcolor{darkslateblue5294141}{blue}, defining all systems with
diamond-shaped graph used for the scatter plot in Figure \ref{fig:conjecture_vgl}\,a).}
\label{tab:conjParamRanges}
\begin{tabular}{|c|c|c|c|c|c|c|c|c|c|c|c|} \hline
& $F_{c1}$ & $F_{c2}$ & $F_{c3}$ & $F_{c4}$ & $F_{a1}$ & $F_{a2}$ & $F_{a3}$ & $F_{a4}$ & $\varphi_{0,2}$ & $\varphi_{0,3}$ & $\varphi_{0,4}$ \\ \hline
{\color{indigo68184}mean} & 1.58 & 3.05 & 1.84 & 1.64 & $-0.01$ & $-0.04$ & 0.05 & 0.11 & 4.39 & 4.79 & $-17.93$ \\ \hline
{\color{indigo68184}variance} & \multicolumn{4}{|c|}{0.5} & \multicolumn{4}{|c|}{0.005} & \multicolumn{3}{|c|}{5} \\ \hline
\hline
{\color{darkslateblue5294141}mean} & 1.25 & 2.75 & 1.24 & 1.34 & $-0.01$ & $-0.27$ & $-0.02$ & $-0.01$ & 2.62 & 3.51 & $-13.29$ \\ \hline
{\color{darkslateblue5294141}variance} & \multicolumn{4}{|c|}{0.5} & \multicolumn{4}{|c|}{2.5} & \multicolumn{3}{|c|}{5} \\ \hline
\end{tabular}
\end{table}
For $j\in\lbrace 1,\dots 4\rbrace$, normally distributed $r_j\sim\mathcal{N}(0,1)$ define
\begin{align}
\omega &= 25.27 + 5\abs{r_1} & & & \text{or} & & & & \omega = 24.67 + 5\abs{r_1}
\end{align}
for the data set plotted in \textcolor{indigo68184}{indigo} and in \textcolor{darkslateblue5294141}{blue}, respectively, and
for both data sets
\begin{align}
n_{\omega,i} = \lfloor 1 + 1.5\abs{r_i}\rfloor.
\end{align} \par
With the exception $k_{13} = 1$ and $k_{31} = \exp\left[-F_1(t)+F_3(t)+ f_{31}\right]$, the transition rates of the
three-state networks used for Figure \ref{fig:conjecture_vgl}\,b) are given by Equation \eqref{Appeq:mgraphtransrates} with
$\kappa_{ij}=1$ and $f_{12}=f_{23}=f_{31}=2$. Moreover, the parameters in the free energies
\begin{align}
F_1(t) &= F_{c1} + F_{a1}\sin{\omega t} \\
F_2(t) &= F_{c2} + F_{a2}\sin(n_\omega\omega t+ \varphi_0) \\
F_3(t) &= F_{c3}
\end{align}
are normally distributed with mean and variance as listed in Table \ref{tab:vglParamRanges}.
\begin{table}
\centering
\caption{Mean and variance of the normal distribution corresponding to the parameters defining the systems used for the
scatter plot in Figure \ref{fig:conjecture_vgl}\,b).}
\label{tab:vglParamRanges}
\begin{tabular}{|c|c|c|c|c|c|c|} \hline
& $F_{c1}$ & $F_{c2}$ & $F_{c3}$ & $F_{a1}$ & $F_{a2}$ & $\varphi_0$ \\ \hline
mean & 1.493 & 0.728 & 1.568 & 0.835 & $-0.349$, $-3.49$ & 23.2 \\ \hline
variance & \multicolumn{5}{|c|}{0.5} & 5 \\ \hline
\end{tabular}
\end{table}
The normally distributed $r_j\sim\mathcal{N}(0,1)$ for $j=1,2$ define
\begin{align}
\omega &= 3.815 + 5\abs{r_1} & & & \text{and} & & & & n_{\omega} = \lfloor 1 + 1.5\abs{r_2}\rfloor.
\end{align} \par
In all cases, we have computed the entropy production rate $\meanpss{\sigma}$ through equation \eqref{eq:setupentropysigma}.
For Figure \ref{fig:conjecture_vgl}\,b), we have also calculated $\pkschaetz$ via Equation \eqref{eq:estimatorpk}, to which
only transitions $(12)$ and $(21)$ contribute. Integrating initial value problems of the absorbing network, where observed
transitions are redirected into auxiliary states \cite{vdm22, seki22}, yields waiting-time distributions
$\wtdpsi{I}{J}{t}{t_0}$. Using the previously obtained probabilities and transition rates, the estimator $\Psischaetz$ can
be determined after integrating out the phase-like time on which all waiting-time distributions depend.
\section{Proof for the entropy estimator \eqref{psiseq:psisEPRPsiEstim}}\label{appen:estimatorpsi}
The observed pairs of directed transitions yield coarse-grained trajectories $\Gamma(t)$
\begin{align}
(I_0,t_{0_0}=0) \overset{\Delta t_1}{\longrightarrow} (I_1, t_{0_1} = t_0+\Delta t_1 \mod \mathcal{T})
\overset{\Delta t_2}{\longrightarrow} (I_2, t_{0_2}) \dots \overset{\Delta t_N}{\longrightarrow}(I_N, t_{0_N})
\end{align}
of length $T$ in time, where we choose the starting time of a period such that we observe the first transition at
$t_{0_0}=0$. A trajectory consists of tuples of observed transition $I_i$ and the phase-like time $t_{0_i}$ of its
observation as well as of waiting times $\Delta t_i$ inbetween. During $\Delta t_i$, an arbitrary number of hidden
transitions can occur. Moreover, as we know $\mathcal{T}$, we know the state of the network directly after each
instantaneous transition. Thus, the next observable transition is independent of the past, i.e., the system has the
Markov property at observable transitions. Therefore, the path weight of coarse-grained trajectories $\Gamma(t)$
factors into
\begin{align}
\pathw{\Gamma(t)} = \mathcal{P}(I_0,0)\wtdpsi{I_0}{I_1}{\Delta t_1}{0}\wtdpsi{I_1}{I_2}{\Delta t_2}{t_{0_1}}
\dots \wtdpsi{I_{N-1}}{I_N}{\Delta t_N}{t_{0_{N-1}}}.\label{Apppsiseq:psisPGammazwei}
\end{align}
This factoring introduces waiting-time distributions of the form
\begin{align}
\wtdpsi{I}{J}{\Delta t}{t_0} \equiv \pathw{\Snippet{I}{J}{\Delta t}{t_0}\middle\vert I,t_0}
= \sum_{\gamma_{I\to J}\in\Gamma_{I\to J}} \mathcal{P}\left[\snippet{I}{J}{\Delta t}{t_0}\middle
\vert I,t_0\right]
\end{align}
defined in Equation \eqref{eq:wtdDefpsismall} in terms of conditional path weights of microscopic trajectories
$\snippet{I}{J}{\Delta t}{t_0}$ that start right after an observed transition $I$ and end with the next observed
transition $J$. \par
The time-reversed process results from reversing the protocol, the trajectory and the time. For a trajectory $\Gamma(t)$ of
length $T$ that starts at the phase-like time $t_0$, the time-reversed transition rates read $\tilde{k}_{ij}(t) =
k_{ij}(T-t)$ while the time transforms as $\tilde{t} = T-t$. Similarly, all quantities obtained by time-reversal will be
marked with a tilde. The path weight of the time-reversed trajectory $\widetilde{\Gamma}$ is, in analogy to Equation
\eqref{Apppsiseq:psisPGammazwei}, given by
\begin{align}
\pathwrev{\widetilde{\Gamma}(t)} &= \widetilde{\mathcal{P}}\left(\tilde{I}_{N+1},T\right)
\wtdpsitrev{\tilde{I}_{N}}{\tilde{I}_{N-1}}{\Delta t_N}{t_{0_{N-1}}+\Delta t_N} \dots
\wtdpsitrev{\tilde{I}_1}{\tilde{I}_0}{\Delta t_1}{0+\Delta t_1}.
\label{Apppsiseq:psisPGammaTildepsieins}
\end{align} \par
Similar to reference \cite{gome08a}, we estimate the entropy production rate $\meanpss{\sigma}$ using the log-sum inequality
(see, e.g., \cite{cove06}) as
\begin{align}
T\meanpss{\sigma} 
&= \sum_{\zeta} \pathw{\zeta(t)} \ln\left(\frac{\pathw{\zeta(t)}}{\pathwrev{\tilde{\zeta}(t)}}\right)\notag \\
&= \sum_{\zeta,\Gamma} \pathw{\Gamma(t)\middle\vert\zeta(t)} \pathw{\zeta(t)} \ln\left(\frac{\pathw{\Gamma(t)
\middle\vert\zeta(t)}\pathw{\zeta(t)}}{\pathwrev{\widetilde{\Gamma}(t)\middle\vert\tilde{\zeta}(t)}
\pathwrev{\tilde{\zeta}(t)}}\right)\notag \\
& \geq \sum_{\zeta,\Gamma}\pathw{\Gamma(t)\middle\vert
\zeta(t)} \pathw{\zeta(t)} \ln\left(\frac{\sum_\zeta\pathw{\Gamma(t)\middle\vert\zeta(t)}\pathw{\zeta(t)}}{
\sum_{\tilde{\zeta}}\pathwrev{\widetilde{\Gamma}(t)\middle\vert\tilde{\zeta}(t)}\pathwrev{\tilde{\zeta}(t)}}\right)
\notag \\    &\equiv T\psischaetz.
\label{Apppsiseq:sigmahatAbschaetzung}
\end{align}
Here, $\pathw{\Gamma(t)\middle\vert\zeta(t)} = 1 =\pathwrev{\widetilde{\Gamma}(t)\middle\vert\tilde{\zeta}(t)}$ holds if
$\Gamma(t)$ is the correct coarse-grained trajectory onto which $\zeta(t)$ is mapped under coarse-graining. Otherwise, these
conditional path weights vanish. \par
Replacing the sums of conditional path weights in the logarithm of the second line of inequality
\eqref{Apppsiseq:sigmahatAbschaetzung} with waiting-time distributions as in Equations \eqref{Apppsiseq:psisPGammazwei} and
\eqref{Apppsiseq:psisPGammaTildepsieins} yields
\begin{align}
&\ln\left(\frac{\pathw{\Gamma(t)}}{\pathwrev{\widetilde{\Gamma}(t)}}\right)
&= \underbrace{\ln\left(\frac{\mathcal{P}\left(I_0,t_{0_0}\right)}{\widetilde{\mathcal{P}}\left(\tilde{I}_{N},T\right)}
\right)}_{\equiv\delta(T,t_{0_0})} + \sum_{j=1}^N \ln\left(\frac{\wtdpsi{I_{j-1}}{I_j}{\Delta t_j}{t_{0_{j-1}}}}{
\wtdpsitrev{\tilde{I}_j}{\tilde{I}_{j-1}}{\Delta t_j}{t_{0_{j}} = t_{0_{j-1}}+\Delta t_j}} \right).
\label{Apppsiseq:psisLogratioPsieins}
\end{align}
The first term on the right hand side, $\delta(T,t_{0_0})$, is periodic when varying one of the fixed times $t_{0_0}$ and
$T$. Hence $\abs{\delta(T,t_{0_0})}\leq c$ holds for a constant $c\in\mathbb{R}^+$. \par
To reformulate the sum on the right hand side of Equation \eqref{Apppsiseq:psisLogratioPsieins}, we define the conditional
counter 
\begin{align}
\nu_{J\vert I}(t,t_0) \equiv \frac{1}{T}\sum_{j=1}^N \delta(t-\Delta t_j)\delta(t_0-t_{0_{j-1}})\delta_{I,I_{j-1}}
\delta_{J,I_j}. \label{Apppsiseq:psiscondcounter}
\end{align}
It sums all terms of trajectories that start with $I$ at $t_0$ and end with the succeeding observable transition $J$ after
waiting time $t$. Substituting the conditional counter into Equation \eqref{Apppsiseq:psisLogratioPsieins} leads to
\begin{align}
\ln\left(\frac{\pathw{\Gamma(t)}}{\pathwrev{\widetilde{\Gamma}(t)}}\right)
=\delta(T,t_{0_0}) + T\int_{0}^{\infty}\int_{0}^{\mathcal{T}} \sum_{I,J} \nu_{J\vert I}(t,t_0)
\ln\left(\frac{\wtdpsi{I}{J}{t}{t_0}}{\wtdpsitrev{\tilde{J}}{\tilde{I}}{t}{t+t_{0}}}\right) \dd{t_0}\dd{t}.
\label{appeq:psiscondcounterTwo}
\end{align} \par
With $\lim_{T\to\infty}\abs{\delta(T,t_{0_0})}/T =0$, the calculation of the expectation value of Equation
\eqref{appeq:psiscondcounterTwo} reduces to determining the expectation value of the conditional counter.
Following Ref. \cite{vdm22}, we argue that
\begin{align}
\nu_{J\vert I}(t,t_0)\Delta t &= \frac{\text{No. of transitions $(IJ)$ per $\mathcal{T}$ after $I$ at $t_0$
and waiting time $t\in[0,\Delta t]$}}{T} \notag \\
&= \frac{\text{No. of transitions $I$ per $\mathcal{T}$ at $t_0$}}{T} P(\text{$J$ after waiting time $t\in[0,
\Delta t]$}\vert I,t_0)
\end{align}
holds true. Together with $n_I(t_0)/\mathcal{T}= \meanpss{n_I} p^\text{pss}(t_0\vert I)$, this results in
\begin{align}
\left\langle \nu_{J\vert I}(t,t_0)\right\rangle &= \sum_\Gamma \nu_{J\vert I}(t,t_0) \pathw{\Gamma(t)} \notag \\ &
= \frac{\left\langle \text{No. of transitions $I$ at $t_0$}
\right\rangle/\mathcal{T}}{T}P(\text{$J$ after waiting time $t\in[0,\Delta t]$}\vert I,t_0) \notag \\
&= \frac{n_I(t_0)}{\mathcal{T}}\wtdpsi{I}{J}{t}{t_0}
= \meanpss{n_I} p^\text{pss}(t_0\vert I)\wtdpsi{I}{J}{t}{t_0}.
\label{Apppsiseq:psisnueexpectation}
\end{align} \par
In total, the estimator $\psischaetz$ of the mean entropy production rate is given by
\begin{align}
\psischaetz &= \left\langle\lim_{T\to\infty}
\frac{1}{T} \ln\left(\frac{\pathw{\Gamma(t)}}{\pathwrev{\widetilde{\Gamma}(t)}}\right) \right\rangle \notag \\ &
= \int_{0}^{\infty}\int_{0}^{\mathcal{T}} \sum_{I,J} \left\langle \nu_{J\vert I}(t,t_0)
\right\rangle \ln\left(\frac{\wtdpsi{I}{J}{t}{t_0}}{\wtdpsitrev{\tilde{J}}{\tilde{I}}{t}{t+t_0}}
\right) \dd{t_0}\dd{t} \notag \\
&= \sum_{I,J} \int_{0}^{\infty}\int_{0}^{\mathcal{T}} \frac{n_I(t_0)}{\mathcal{T}}\wtdpsi{I}{J}{t}{t_0}
\ln\left(\frac{\wtdpsi{I}{J}{t}{t_0}}{\wtdpsitrev{\tilde{J}}{\tilde{I}}{t}{t+t_0}}\right) \dd{t_0}\dd{t},
\label{Apppsiseq:psisEPRPsiEstim}
\end{align}
where the second equality follows from the vanishing $\delta(T,t_{0_0})/T$ in the limit $T\to\infty$. The estimator is
non-negative as its definition \eqref{Apppsiseq:sigmahatAbschaetzung} has the form of a Kullback-Leibler divergence. In
the special case of a NESS, rewriting $\psischaetz$ using Equation \eqref{Apppsiseq:psisnueexpectation} reveals that
this estimator reduces to $\Psischaetz$, which we define in Equation \eqref{psiseq:PsisEPRPsiEstimconj}, since then
$p^\text{pss}(t_0\vert I)=1/\mathcal{T}$ and the waiting-time distributions do not depend on $t_0$. \par
\end{appendices}

\end{document}